\documentclass[final,5p,times,twocolumn]{elsarticle}
\usepackage{epsfig}
\usepackage{amssymb}
\usepackage{amsthm}
\biboptions{comma,square}
\journal{Journal of Magnetism and Magnetic Materials}
\begin{document}
\begin{frontmatter}
\title{Exchange-induced phase separation in Ni-Cu films}
\author[focal]{A.F.~Kravets\corref{cor}}
\ead{anatolii@kth.se}
\cortext[cor]{Corresponding author}
\author[focal]{A.N.~Timoshevskii}
\author[focal]{B.Z.~Yanchitsky}
\author[focal]{O.Yu.~Salyuk}
\author[focal]{S.O.~Yablonovskii}
\author[els]{S.~Andersson}
\author[els]{V.~Korenivski}
\address[focal]{Institute of Magnetism, National Academy of Sciences of Ukraine, Vernadsky 36 b, 03142 Kyiv, Ukraine}
\address[els]{Nanostructure Physics, Royal Institute of Technology,
10691 Stockholm, Sweden}

\title{Exchange-induced phase separation in Ni-Cu films}

\begin{abstract}
Magneto-structural properties of films of diluted ferromagnetic
alloys Ni$_x$Cu$_{1-x}$ in the concentration range $0.7 < x < 1.0$
are studied experimentally. Films deposited by magnetron sputtering
show partial phase separation, as evidenced by structural analysis
and ferromagnetic resonance measurements. The phase diagram of the
Ni$_x$Cu$_{1-x}$ bulk system is obtained using numerical theoretical
analysis of the electronic structure, taking into account the
inter-atomic exchange interactions. The results confirm the
experimentally found partial phase separation, explain it as
magnetic in origin, and indicate an additional metastable region
connected with the ferromagnetic transition in the system.
\end{abstract}

\end{frontmatter}
\section{Introduction}
Magnetic multilayered films have been receiving much attention from
the research community in the field of spintronics. Recently, a
device has been proposed \cite{Kadigrobov_2010JAP} and demonstrated
\cite{Andersson_2010JAP, Andersson_2010ITM}, in which the magnetic
and spin-transport properties are controlled thermo-electrically. At
the core of the new design is a tri-layer (\textit{F/f/F}) of two
strong ferromagnets (\textit{F}) exchange-coupled by a weakly
ferromagnetic spacer (\textit{f}) having the Curie point at/or near
room temperature ($300-500$~K). The spacer~\textit{f}, controlling the exchange coupling between the outer layers, can be made of a diluted ferromagnetic alloy, such as Ni$_x$Cu$_{1-x}$ \cite{Andersson_2010JAP, Andersson_2010ITM}. It is desirable that the spacer is compositionally uniform and does not contain multiple phases, which can form as a result of atomic segregation during film deposition or subsequent heat treatment.

Alloys Ni$_x$Cu$_{1-x}$ in their bulk form are \textit{fcc} binary
substitution alloys ($\alpha$-phase), with Ni and Cu mutually
solvable in any proportion up to the temperature of 627~K, where
at $x=0.67$, the $\alpha$ phase separates into two, $\alpha_1$ and
$\alpha_2$ \cite{Chakrabarti_1994ASM}. According to
\cite{Chakrabarti_1994ASM}, the phase diagram contains a metastable
region with phase separation at the nickel concentration of
$0.86<x<0.91$, which can lead to additional magneto-structural
inhomogeneities in the system. The results, obtained in this work,
however, are obtained using an empirical model, based on the
experimental data. It is desirable to independently simulate the
phase equilibria in the system Ni$_x$Cu$_{1-x}$, first without using
experimental information and then compare the results with the
experimental findings. Of particular interest is the influence of
the magnetic interactions in the system on the potential structural
and/or compositional phase separation. The Curie temperature of bulk
Ni$_x$Cu$_{1-x}$ alloys depends linearly on the Ni concentration
\cite{Chakrabarti_1994ASM}. By varying the Ni content from 0.5 to 1
the Curie temperature is changed from 0 K to 627 K. However, based
on neutron scattering data, it was shown that both bulk samples
\cite{Hicks_1969PRL,Robbins_1969PRL} and sputtered films of Ni-Cu
alloys  \cite{Iannone_2007PRB} exhibit partial phase separation and
form Ni rich clusters of typical size $5-10$~\AA{} and
magnetic moment  $8-12$~$\mu_B$. The formation of such Ni
clusters and the resulting magnetic inhomogeneity in the material
can lead to anomalous magnetic \cite{Kidron_1970PhLA} and transport
\cite{Houghton_1970PRL} properties.

In this work we investigate the mechanisms behind phase formations
in diluted ferromagnetic alloy films of Ni$_x$Cu$_{1-x}$ and explain
using first-principles calculations the enhancing effect of the
exchange interaction on phase separation in the ferromagnetic
composition range of $0.7 < x < 1.0$. We discuss optimum ways to
prepare films with enhanced magneto-structural homogeneity.

\section{Experimental methods}
Films of diluted ferromagnetic Ni$_x$Cu$_{1-x}$ alloys, with
$0.7<x<0.9$ in Ni concentration, 100 nm thick, were deposited at
room temperature on thermally oxidized Si substrates using DC
magnetron co-sputtering from Cu and Ni targets. Substrates of
$150\times10$ mm in size, placed above the Cu and Ni targets, along
the line connecting the centers of the targets, were used to produce a continuous and essentially linear compositional gradient along the substrate strip. The base pressure in the deposition chamber was $\sim
5\times10^{-8}$ Torr and the Ar pressure used during deposition was
$5$ mTorr. The deposition rate for Ni and Cu in the center of the
substrate was $\sim 0.5$~\AA{}/sec. Samples for magneto-structural
measurements were cut from the substrate strip into short sections  with dimensions of $5\times10$ mm. In total, 30 samples of
various composition Ni$_x$Cu$_{1-x}$ were produced in the same
fabrication cycle under the same conditions. The thickness of the
films was determined using a surface profilometer.
The composition of the films was determined using x-ray dispersion
spectroscopy analysis.

Ferromagnetic resonance (FMR) was used to determine the effective
magnetization of the Ni$_x$Cu$_{1-x}$ films ($M_{eff}$) and their
Curie temperature ($T_c$). FMR measurements were performed at $9.45$
GHz using a Bruker ELEXSYS-E500 spectrometer equipped with a
goniometer for angle-dependent measurements and a variable
temperature cryostat. The effective magnetization was obtained from
the resonance fields using the Kittel formulae \cite{Kittel_1948PR}:
\begin{eqnarray}\label{FMREq}
\frac{\omega}{\gamma} &=& H_{\perp} - 4 \pi M_{eff} \nonumber\\
\frac{\omega}{\gamma} &=& \sqrt{H_{\parallel} \left(H_{\parallel} +
4 \pi M_{eff} \right) },
\end{eqnarray}
where $H_{\perp}$ and $H_{\parallel}$ are the measured perpendicular
and in-plane FMR resonance fields, respectively,  $\omega$ --
frequency, and  $\gamma$ -- the gyromagnetic ratio.

The Curie temperature of the films was determined from the FMR data
as temperature at which either $M_{eff}$ or the FMR signal vanished. The degree of magneto-structural non-uniformity in the films was estimated from the width of the FMR peaks (magnetic non-uniformity leads to broader FMR peaks).

Mechanical stress arising from mismatches in the lattice parameters
of the substrate and the films can lead to perpendicular to the plane magnetic anisotropy in the films through magnetostriction. This magnetostrictive contribution to $M_{eff}$ can be estimated using 
\cite{Bozorth_1993}:
\begin{equation}\label{FMREqH}
4 \pi M_{eff} = 4 \pi M_{s} - H_{\perp}^\ast.
\end{equation}
Here $M_s$ is the saturation magnetization of the film,
$H_{\perp}^\ast = -2\lambda\sigma/3$ -- magnetostriction-induced
perpendicular-anisotropy field, $\lambda$ -- magnetostrictive
constant, $\sigma$ -- mechanical stress in the film. It follows from
(\ref{FMREqH}) that for a strong magnetostrictive contribution
and/or small saturation magnetization, the effective magnetization
can become negative, which corresponds to its
perpendicular-to-the-plane orientation in the absence of external
fields~\cite{Kittel_1948PR}.

\section{Experimental results and discussion}
\begin{figure}[!htb]
\begin{center}
\includegraphics[scale=0.95]{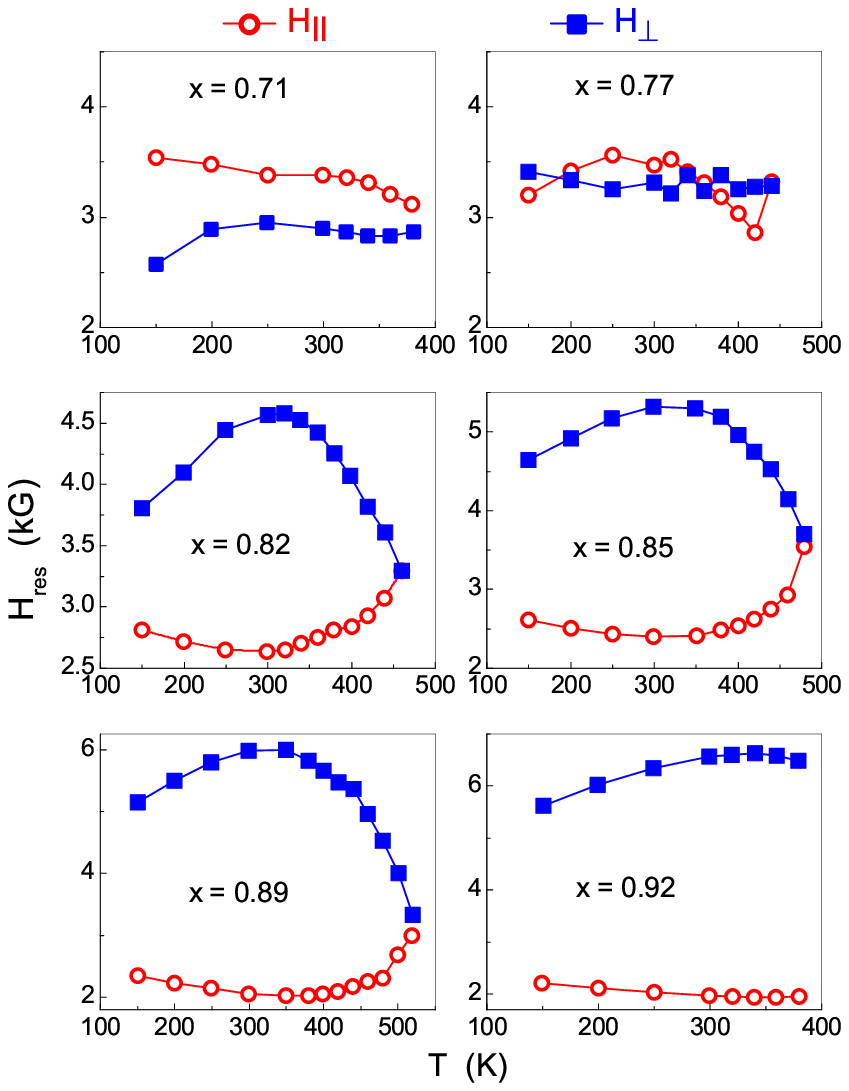}
\end{center}
\caption[]{ FMR resonance fields for Ni$_x$Cu$_{1-x}$ films versus
temperature.
(\raisebox{-0.5mm}{\includegraphics[scale=0.36]{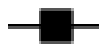}}) --
external field perpendicular to the film plane,
(\raisebox{-0.5mm}{\includegraphics[scale=0.36]{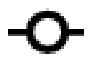}}) -- field
in the film plane. }
\label{Hres(T)} 
\end{figure}

Figure \ref{Hres(T)} shows the temperature dependence of the FMR
resonance fields for Ni$_x$Cu$_{1-x}$ films with different Ni
content. For Ni$_{0.71}$Cu$_{0.29}$ and Ni$_{0.77}$Cu$_{0.23}$ films,  $H_{\perp} < H_{\parallel}$ in a broad temperature range, which
indicates an out-of-plane magnetization orientation for these compositions. For films with higher concentrations of Ni ($x=0.82-0.92$), $H_{\perp} > H_{\parallel}$ and the magnetization is in-plane at zero field.

\begin{figure}[!htb]
\begin{center}
\includegraphics[scale=0.95]{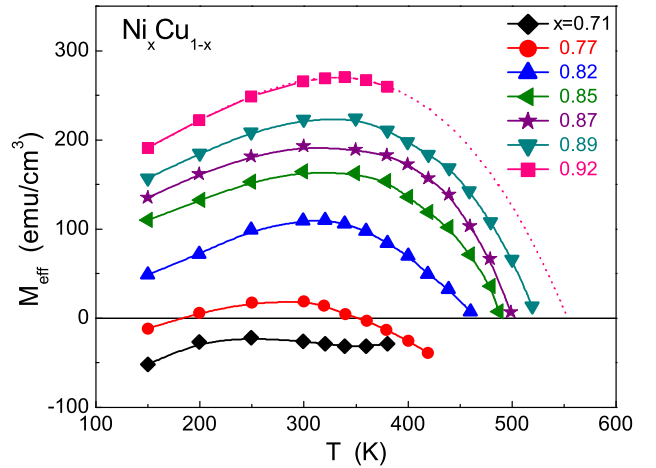}
\end{center}
\caption[]{ Temperature dependence of the effective magnetization of
Ni$_x$Cu$_{1-x}$ films with different Ni content $x$. Dashed line is
the approximation for $x=0.92$.}
\label{Meff(T)} 
\end{figure}

The temperature dependence of the effective magnetization obtained
using (\ref{FMREq}) and the measured resonance fields (see~Fig.~\ref{Hres(T)}) are shown in Fig.~\ref{Meff(T)}. For all
compositions, the effective magnetization first increases with
increasing temperature from $150$ K to $\sim320$ K and subsequently
decreases above $T > 320$~K. This indicates a significant
magnetostrictive contribution to the anisotropy. For low Ni
concentrations, Ni$_{0.71}$Cu$_{0.29}$ and Ni$_{0.77}$Cu$_{0.23}$,
$M_{eff}$ is negative, while for Ni$_{0.82}$Cu$_{0.18}$,
Ni$_{0.85}$Cu$_{0.15}$, Ni$_{0.87}$Cu$_{0.13}$,
Ni$_{0.89}$Cu$_{0.11}$ and Ni$_{0.92}$Cu$_{0.08}$ $M_{eff} > 0$.
Thus, in films with low Ni concentrations and therefore low
saturation magnetization, magnetostriction results in perpendicular
magnetic anisotropy.

\begin{figure}[!htb]
\begin{center}
\includegraphics[scale=0.95]{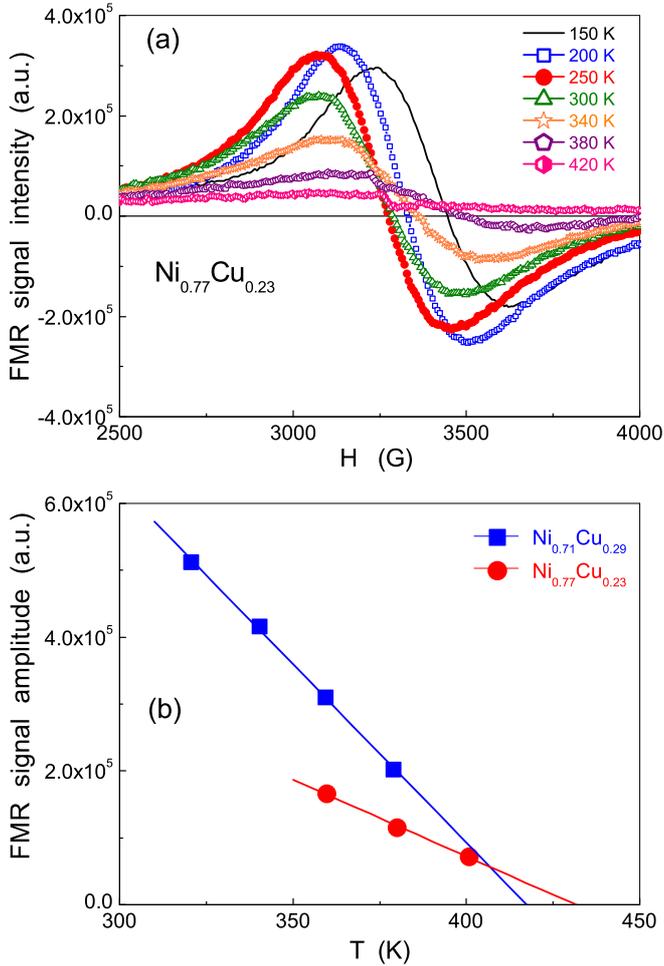}
\end{center}
\caption[]{ A set of FMR lines for different temperatures for
Ni$_{0.77}$Cu$_{0.23}$, measured with the in-plane field (a). FMR
signal amplitude for the in-plane field configuration for
Ni$_{0.71}$Cu$_{0.29}$ and Ni$_{0.77}$Cu$_{0.23}$ films
(b).}
\label{FMRlines} 
\end{figure}

The temperature of the Curie transition from the ferromagnetic to
the paramagnetic state ($T_c$) for Ni$_x$Cu$_{1-x}$ films with
$x=0.82$; 0.85; 0.87; 0.89 and 0.92 was determined as the
temperature at which $M_{eff} \rightarrow 0$. For
Ni$_{0.71}$Cu$_{0.29}$ and Ni$_{0.77}$Cu$_{0.23}$ films with strong
magnetostriction and out-of-plane magnetization, it was more
appropriate to determine the $T_c$ directly from the temperature
dependence of the respective amplitudes of the FMR signal
(Fig.~\ref{FMRlines}). $T_c$ in this case was determined as the
temperature at which the FMR signal amplitude approached zero
(Fig.~\ref{FMRlines}).

\begin{figure}[!htb]
\begin{center}
\includegraphics[scale=0.71]{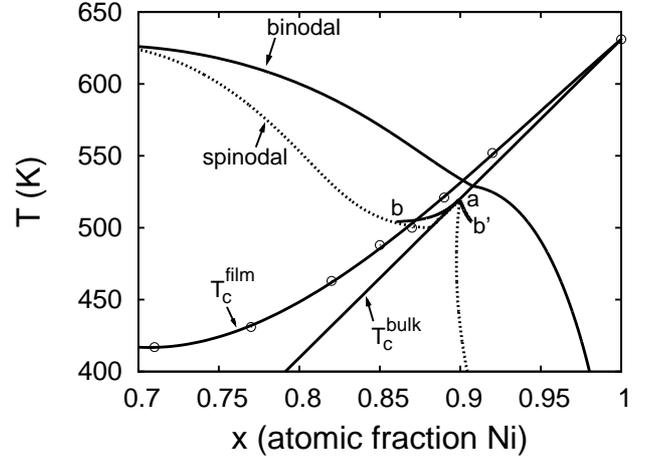}
\end{center}
\caption[]{ Phase diagram according to \cite{Chakrabarti_1994ASM}
and the Curie temperature versus concentration for bulk
Ni$_x$Cu$_{1-x}$ ($T_c^{\textrm{bulk}}$) \cite{Chakrabarti_1994ASM}
and thin films ($T_c^{\textrm{film}}$,
\raisebox{-0.5mm}{\includegraphics[scale=0.4]{circle.eps}}, this
work). Lines $ab-ab'$ correspond to metastable equilibria. }
\label{PhaseDiag1} 
\end{figure}

$T_c$ as a function of concentration in Ni$_x$Cu$_{1-x}$ films,
determined from the FMR data as described above, is shown in
Fig.~\ref{PhaseDiag1}. The figure also shows the phase diagram
and $T_c$ for the same system in the bulk form
\cite{Chakrabarti_1994ASM}. The phase diagram also shows the
lines of the binodal, spinodal, as well as the metastable regions,
the existence of which was predicted in \cite{Chakrabarti_1994ASM}.
Fig.~\ref{PhaseDiag1} shows that in the vicinity of the
metastable equilibria (lines $ab-ab'$), the concentration
dependence of the $T_c$ for bulk \cite{Chakrabarti_1994ASM} and film
materials (this work) are significantly different. In films,
the behaviour of $T_c(x)$ is significantly non-linear, while in the
bulk it is practically linear.

The non-linear $T_c$ vs. x in films indicates that magnetic and
structural inhomogeneities increase with increasing Cu content,
specifically in the concentration range $0.7 < x < 0.9$. The fact
that $T_c$ in films is always higher than in the bulk
(Fig.~\ref{PhaseDiag1}) suggests that in sputter deposited
films regions or clusters rich in Ni compared to the nominal
composition of the film are formed. The $T_c$ of such Ni-rich regions
is higher than the $T_c$ for the nominal composition.

\begin{figure}[htb]
\begin{center}
\includegraphics[scale=0.95]{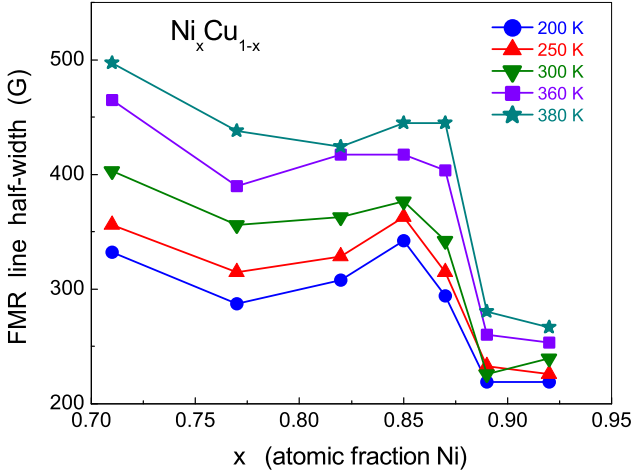}
\end{center}
\caption[]{ FMR line width versus concentration for Ni$_x$Cu$_{1-x}$
films, measured at different temperatures. }
\label{FMRLineWidth} 
\end{figure}

Additional information about the non-uniformities in the films can
be obtained from the analysis of the FMR line width.
Fig.~\ref{FMRLineWidth} shows the half width of the FMR lines for
Ni$_x$Cu$_{1-x}$ films, for various temperatures
between 200 and 380~K. Judging from the relatively narrow line
widths (Fig.~\ref{FMRLineWidth}), the films can be considered
rather uniform magnetically, at least within the ferromagnetic
exchange length of $\sim 10 $~nm. It is also seen from the
data that the degree of non-uniformity depends on the concentration
in Ni$_x$Cu$_{1-x}$ films. Thus, at small Cu concentrations, $0.89 <
x < 1.0$, the alloy has a narrow FMR line indicating relatively good
magneto-structural uniformity. For higher Cu concentrations, $0.70 <
x < 0.89$, the FMR line broadens, which indicates a higher magnetic
non-uniformity. This, combined with the observed enhanced $T_c$ from
Ni clustering, indicates that in this interesting concentration
range near the metastable region of the phase diagram the films can
become structurally non-uniform, possibly consisting from multiple
compositional phases. Below we examine theoretically the microscopic mechanism of phase separation in the system, taking into account the
interatomic magnetic interactions.

\section{Calculation of phase equilibria in
N\lowercase{i}$_x$C\lowercase{u}$_{1-x}$ system}
There are no direct experimental confirmations that magnetic
interactions in Ni$_x$Cu$_{1-x}$ could be responsible for additional
phase equilibria in the system. Temperatures required for reaching
equilibrium conditions are difficult to attain experimentally, if at
all possible. Nevertheless, it was predicted
\cite{Chakrabarti_1994ASM} that in Ni$_x$Cu$_{1-x}$ alloys with
$(0.86<x<0.91)$ a metastable miscibility gap should exist due to
magnetic interactions. For describing the magnetic contribution to
the free energy, the authors of \cite{Chakrabarti_1994ASM} used a
model proposed by Chuang \cite{Chuang_1985MTA,Chuang_1986MTA}. The
model contains an empirical expression for the heat capacity of a
ferromagnetic alloy, with adjustable parameters obtained from the experiment. In order to obtain a deeper insight into whether such
miscibility gap could indeed exist and be due to magnetic
interactions, it is highly desirable to first construct a solid
state phase diagram of Ni$_x$Cu$_{1-x}$ system using
experiment-independent first-principle calculations of interatomic
interactions, and then compare the results with the experiment.

For describing the Ni$_x$Cu$_{1-x}$ alloy we use a model Hamiltonian
taking into account the Ni-Ni exchange interactions.  The
Hamiltonian is similar to the one used in \cite{Taylor_1992JMM} for
describing Fe$_x$Ni$_{1-x}$ alloys. The approach relies on a lattice
model, with each site $i$ of the \textit{fcc} lattice being occupied
either by an atom of Ni ($c_i=1$) or Cu ($c_i=0$). The atomic
fraction of Ni ($x$) will be taken as $c$. The magnetic interactions
between the Ni atoms in the system are described by the classical
isotropic Heisenberg model. Since the local magnetic moment of Cu
atoms is negligible, the magnetic interactions of Cu-Cu and Cu-Ni
are omitted. Thus, the model Hamiltonian of the system is:
\begin{equation}\label{EqCE}
E=\frac{1}{2} \sum_{ij} v_{ij} c_i c_j + \frac{1}{2} \sum_{ij}
J_{ij} c_i c_j \mathbf{s}_i \mathbf{s}_j,
\end{equation}
where $v_{ij}$ are the mixing potentials responsible for "chemical"
interactions between the atoms, $J_{ij}$ -- magnetic exchange
constants for Ni-Ni interactions (independent of concentration $c$),
$\mathbf{s}_i$ -- vector of unit length along the direction of the
Ni local magnetic moment. Values of unknowns $v_{ij}$ for first 4
coordination shells and $J_{ij}$ for 2 shells were obtained as
solutions of a  system of equations (\ref{EqCE}) for a set of
ordered superstructures representing the Ni$_x$Cu$_{1-x}$ alloys.
The structures have various distributions of nickel atoms and
different magnetic ordering. Since the most interesting region of
the phase diagram is at high concentrations of Ni, 5 superstructures
of composition CuNi$_7$ (32 atoms per unit cell, ferromagnetic
ordering) and 3 structures of pure Ni with ferromagnetic and
antiferromagnetic ordering were selected.

The total free energy of the structures were calculated within the
density functional theory (DFT) using the FLAPW method (Wien2k
package \cite{Wien2k}). The calculations were performed taking into
account spin polarization, and the GGA exchange correlation potential
was taken to be of the form of \cite{Perdew_1996PRL}. The MT-radii
for Cu and Ni were equal to 2.2~\textit{a.u.}, the electron density
was calculated using 500~k-points in the first Brillouin zone.
Structural relaxation was performed for the lattice parameters and
the atomic positions. The atomic positions were iterated until the
forces on the nuclei became smaller than 1~mRy/a.u. The accuracy of the total energy was approximately 1~meV. The calculated values of the
interatomic interactions (in meV) were: $v_{ij}=\{\textrm{-16.37,
32.55, -7.72, -2.52}\}$, $J_{ij}=\{\textrm{-6.78, -18.27}\}$. The
calculation of the phase equilibria was performed within the
mean-field approximation for the Hamiltonian of (\ref{EqCE}). All Ni
atoms were taken to have the same value of magnetic moment
$\mathbf{m}=(0,0,m)$. $m$ is obtained as \cite{BuschowBoer_2003}
$m=L(H/k_BT)$, with $H=-(\displaystyle \sum \limits_j J_{1j} ) m c$
being the effective magnetic field, $k_B$ -- Boltzmann constant,
$T$ -- absolute temperature, $L(x)$ -- Langevin function. The free
energy is given by the expression:
\begin{eqnarray}\label{EqFreeEnrg}
F  & = & \frac{1}{2} \left( \sum_{j} v_{1j} \right) c^2 +
 k_B T \left( c \ln c + (1-c) \ln(1-c) \right) \nonumber\\
& {} & + \frac{1}{2} H m c - k_B T c \ln \frac{\sinh(H/k_B
T)}{H/k_BT}.
\end{eqnarray}

\begin{figure}[!htb]
\begin{center}
\includegraphics[scale=0.71]{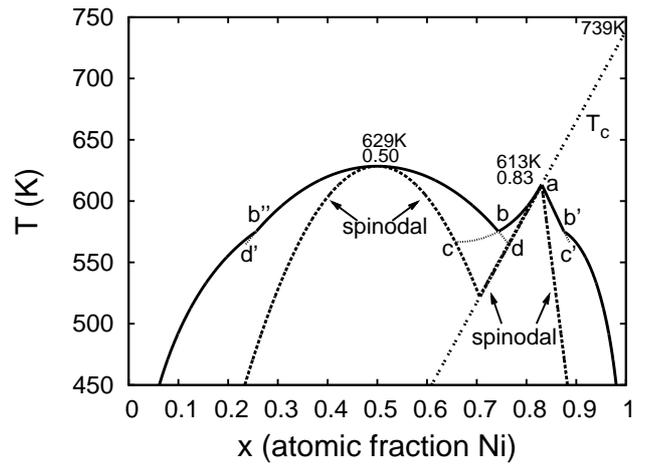}
\end{center}
\caption{Calculated phase diagram for Ni$_x$Cu$_{1-x}$ (crystalline
state).}
\label{PhaseDiag2}
\end{figure}

Phase equilibria were calculated using the standard ``common
tangent'' method. The obtained phase diagram of the solid
Ni$_x$Cu$_{1-x}$ is shown in Fig. \ref{PhaseDiag2}. The lines
shown are the binodal, spinodal, metastable equilibria, and the Curie
temperature $T_c$ of the alloy versus composition. The Curie
temperature below the binodal line corresponds to a homogeneous
solid solution. Metastable equilibria $b''d'-bd$ and
$bc-b'c'$ are shown as dashed lines.

The phase diagram contains a stable miscibility gap (lines
$ab-ab'$) due to magnetic transformations in the alloy. A
similar miscibility gap exists in the Fe-Ni system (phases
$\gamma_1$, $\gamma_2$ \cite{Chuang_1986MTA}). The presence of this
miscibility gap in Ni$_{x}$Cu$_{1-x}$ system results in additional
equilibria of phases with close compositions but different magnetic
properties and, importantly, different Curie temperatures. The
presence of different phases in the alloy should broaden the
magnetic transition, which is indeed observed in our
Ni$_{x}$Cu$_{1-x}$ films.

Possibly due to our Ni atomic magnetic moments taken to be
independent of composition, the obtained theoretical $T_c$ versus
composition (Fig.~\ref{PhaseDiag2}) deviates from the measured one
(Fig.~\ref{PhaseDiag1}). It should also be mentioned, that in
contrast to the results of \cite{Chakrabarti_1994ASM}, our
calculations show that the additional miscibility gap is stable.

Thus, our ab-initio numerical theoretical analysis, not relying on
experimental data, shows additional phase equilibria in the
Ni$_x$Cu$_{1-x}$ system at high concentrations of Ni. These
additional equilibria are due to the ferromagnetic interactions in
the system.

Using the expression for the free energy of Eq.
(\ref{EqFreeEnrg}) it is possible to obtain the general condition
for the existence of an additional miscibility gap. For this, it is
convenient to introduce the following quantities: $V_0=\sum_{j}
v_{1j}$, $J_0=\sum_{j} J_{1j}$. The lines of the spinodal and magnetic
transformations intersect at point \textit{a}
(Fig.~\ref{PhaseDiag2}), which gives two conditions: $\partial^2
F/\partial^2 c=0$, $T=T_c$.

The expressions for the spontaneous magnetization near $T_c$,
$T<Tc$: $m=\sqrt{5/3} \sqrt{(T_c-T)/T_c}$, and that for the Curie
temperature $T_c=-J_0 c /3k$, yield the following equation for the
concentration at point \textit{a}:
\begin{eqnarray}
\frac{\partial^2 F}{\partial^2 c} = V_0 - \frac{1}{3(1-c_a)} J_0 +
\frac{5}{6} J_0 = 0.
\end{eqnarray}
Thus:
\begin{eqnarray}
c_a = \frac{6 + 3 J_0/V_0}{6 + 5 J_0/V_0}.
\end{eqnarray}
Condition $0 < c_a < 1$ yields for the interaction parameters:
$J_0/V_0 > 0$, This in turn gives the concentration range for the
miscibility gap: $3/5 < c_a < 1$. To sum up, for binary alloys with
decomposition ($V_0 < 0$) and with one element being ferromagnetic
($J_0 < 0$), there must exist an additional miscibility gap due to
the inter-atomic magnetic interactions.

The obtained phase diagram indicates that the magnetic and
structural properties of the samples should strongly depend on the
temperature of  fabrication (substrate temperature for films).
Specifically, quenching the material on to substrates is expected to
be beneficial for the uniformity of the films. Here, a uniform
mixing of the two elements in the magnetron plasma flux is frozen in
on the substrate before the magnetic atoms have time to interact by
exchange. If annealed at a sufficiently high temperature and then cooled slowly, the exchange interaction in the film is expected to result in atomic diffusion as to favor Ni-clustering and therefore partial phase separation into a Ni-rich and Cu-rich phases, as discussed above.

\section{Conclusions}
Diluted ferromagnetic alloy films Ni$_x$Cu$_{1-x}$ obtained by
magnetron sputtering exhibit partial phase separation into
Ni-rich and Cu-rich phases. The phase separation is more pronounced for higher Cu dilution and leads to a broadening of the ferro- to
paramagnetic phase transition in the system. This process is
modeled using ab-initio calculations, taking into account the
magnetic interactions in the system. Our analysis shows that the
phase separation in the concentration range of interest is due to
the Ni-Ni exchange interaction, which favors clustering of Ni and
thereby compositional gradients in the alloy. This can additionally
lead to a significantly non-linear dependence of the Curie
temperature on the alloy concentration. Our analysis further
suggests that the general nature of the observed phase separation in
Ni-Cu, namely the exchange-induced magnetic atom clustering in an
otherwise perfectly solvable two-component system, should be found
in other diluted alloys of magnetic-nonmagnetic or strongly
magnetic-weakly magnetic elements and likewise lead to
magneto-structural inhomogeneities there.

\section*{Acknowledgements}
Authors would like to thank the FP7-FET-STELE project for providing
financial support for this study.


%
\end{document}